\documentclass[preprint]{aastex7}

\usepackage{amsmath}
\usepackage{hyperref}
\usepackage{siunitx}
\DeclareSIUnit{\cm}{\centi\metre}
\DeclareSIUnit{\erg}{erg}
\DeclareSIUnit{\ergs}{\erg\per\second}
\DeclareSIUnit{\PeV}{\peta\eV}
\DeclareSIUnit{\clight}{\mathit{c}}
\usepackage{booktabs}
\usepackage{threeparttable}

\begin{document}

\title{Microquasar Cygnus X-3 as the PeVatron powering the Cygnus Bubble}

\author[0009-0009-1182-341X,gname=Zhaodong, sname=Shi]{Zhaodong Shi}
\affiliation{Department of Astronomy, School of Physical Sciences,
University of Science and Technology of China, Hefei 230026, Anhui, China}
\affiliation{School of Astronomy and Space Science, University of Science and Technology of China, Hefei 230026, Anhui, China}
\email[show]{shizd@ustc.edu.cn}

\author[gname=Guangwei, sname=Wang]{Guangwei Wang}
\affiliation{Department of Astronomy, School of Physical Sciences,
University of Science and Technology of China, Hefei 230026, Anhui, China}
\affiliation{School of Astronomy and Space Science, University of Science and Technology of China, Hefei 230026, Anhui, China}
\email[show]{wangguangwei@mail.ustc.edu.cn}

\author[orcid=0000-0001-5801-2547,gname='Rui-zhi', sname=Yang]{Rui-zhi Yang} 
\affiliation{Department of Astronomy, School of Physical Sciences,
University of Science and Technology of China, Hefei 230026, Anhui, China}
\affiliation{School of Astronomy and Space Science, University of Science and Technology of China, Hefei 230026, Anhui, China}
\affiliation{TIANFU Cosmic Ray Research Center, Chengdu 610000, Sichuan, China}
\email[show]{yangrz@ustc.edu.cn}

\author[orcid=0000-0003-1157-3915,gname='Felix', sname=Aharonian]{Felix Aharonian}
\affiliation{University of Science and Technology of China, Hefei 230026, Anhui, China}
\affiliation{Yerevan State University, 1 Alek Manukyan Street, Yerevan 0025, Armenia}
\affiliation{TIANFU Cosmic Ray Research Center, Chengdu 610000, Sichuan, China}
\affiliation{Max-Planck-Institut for Nuclear Physics, P.O. Box 103980, 69029 Heidelberg, Germany}
\email[show]{felix.aharonian@mpi-hd.mpg.de}

\begin{abstract}
The recent discovery by the LHAASO collaboration of a variable ultra-high-energy (UHE; $E_\gamma \ge \qty{100}{\TeV}$) $\gamma$-ray source associated with the microquasar Cygnus X-3, with a spectrum extending to several PeV, provides compelling evidence for a hadronic super-PeVatron operating within the binary system. Inside the binary, the accelerated protons lose only a small fraction of their energy; upon escaping into the interstellar medium, they propagate diffusively to form a vast gamma-ray ``halo” structure extended to hundreds of parsecs. We argue that this halo has already been detected and corresponds to the Cygnus Bubble, an extended UHE $\gamma$-ray source reported by the LHAASO collaboration -- which possesses an angular extension of $\approx \qty{6}{\degree}$ and an energy spectrum reaching 1 PeV. While the Cygnus Bubble is generally attributed to the star-forming region Cygnus X (specifically the Cygnus OB2 association at \qty{1.4}{kpc}), we demonstrate that an association with Cygnus X-3 is physically more natural at energies above \qty{400}{\TeV}. This is supported by the cosmic-ray radial distribution, derived from the $\gamma$-ray and gas distributions, which points to continuous injection from a point-like source. The energetic requirements of the central accelerator are reasonably affordable and feasible. This reassignment identifies the Cygnus Bubble as a member of the recently discovered population of microquasar UHE $\gamma$-ray halos.

\end{abstract}



\section{Introduction} \label{sec:intro}
Recent detections of dozens of ultra-high-energy (UHE) $\gamma$-ray sources \citep{Chen2024} have revealed a growing population of Galactic PeVatrons ---  cosmic accelerators capable of boosting particles to PeV energies. Most of these sources are spatially extended $\gamma$-ray structures associated with pulsar wind nebulae (PWNe), molecular clouds in the proximity of middle-aged supernova remnants (SNRs), and bubbles surrounding microquasars, hereafter referred to as ``microquasar halos" ($\mu$QSHs). Generally, the $\gamma$-ray emission does not spatially coincide with the particle accelerator itself; rather, the $\gamma$-ray morphology reflects the complex interplay between the spatial distribution of relativistic particles and the ambient target medium. Consequently, identifying the specific acceleration sites remains challenging and requires deep phenomenological and theoretical studies based on comprehensive multi-wavelength analysis.

In $\mu$QSHs, relativistic particles may originate within the compact binary system, at the parsec-scale termination of the outflow (jets or winds), or both. For most sources, current observations cannot distinguish between these sites. Cygnus X-3, however, stands out as a unique case; the reported UHE $\gamma$-ray variability \citep{Cao2025}, specifically its modulation at the \qty{4.8}{\hour} orbital period, indicates that acceleration and emission occur within a compact region, likely associated with a (sub)relativistic wind or jet inside the binary. The production of UHE $\gamma$ rays extended beyond \qty{1}{\PeV} inside the binary system points to the presence of a hadronic SuperPeVatron, precluding any ``leptonic" contribution; because of severe synchrotron and inverse Compton losses, even at the maximum possible theoretical acceleration rate, electrons cannot reach PeV energies \citep{Cao2025}. Notably, a hadronic PeVatron localized within the binary system does not preclude the further acceleration of PeV particles in a large-scale ``jet-termination" scenario.

Within the binary environment, the accelerated protons lose only a small fraction of their energy and escape into the surrounding medium without significant deformation of their initial (acceleration) spectrum. Upon entering the interstellar medium (ISM), these particles propagate diffusively, interacting with the ambient gas to form a vast $\gamma$-ray ``halo" extending over hundreds of parsecs. The detectability of such a structure depends primarily on the medium's diffusion coefficient and the target gas density. Generally, in the vicinity of the source, the cosmic-ray diffusion coefficient must be substantially smaller than the ``standard" Galactic value to prevent the rapid runaway of protons and maintain an adequate surface brightness distribution. Such an enhancement of diffusive confinement is physically expected in the surroundings of powerful cosmic-ray accelerators \citep{Malkov2013,DAngelo2016,Nava2016,Nava2019,Schroer2022}. Finally, the detection of these extended features requires instruments highly sensitive to low-surface-brightness emission; in this regard, the LHAASO KM2A detector has uniquely demonstrated its potential to reveal diffuse UHE $\gamma$-ray sources throughout the Galactic plane.

Notably, it is likely that a halo around Cygnus X-3 has already been detected. Indeed, the Cygnus Bubble, reported by the LHAASO collaboration as an extended ultra-high-energy (UHE) $\gamma$-ray source with an angular radius of $\approx \qty{6}{\degree}$ toward the Cygnus region \citep{Cao2024}, matches this description. This statement, at first glance, may seem paradoxical as it opposes the widely accepted view that the Cygnus Bubble is associated with the Cygnus X star-forming region, specifically the Cyg OB2 cluster at $d_1=\qty{1.4}{kpc}$ \citep{Aharonian2019,Menchiari2024,Harer2025}. One might argue that assuming the larger distance of Cygnus X-3 ($d_2 \approx \qty{9.7}{kpc}$) increases the required $\gamma$-ray luminosity by $(d_2/d_1)^2$, or roughly 1.5 orders of magnitude.

However, this does not present a fundamental problem. Depending on the radial dependence of the product of the cosmic ray (CR) and gas densities, a distant source can be energetically competitive. The observed $\gamma$-ray flux, $F_\gamma$, resulting from the interaction of CRs with the ambient gas, is proportional to:

\begin{equation} \label{eq:gamfluxsph}
    F_\gamma \propto \frac{1}{d^2} \int_0^R N_{\mathrm{CR}}(r) n_\mathrm{g}(r) 4\pi r^2\,dr.
\end{equation}

\noindent Assuming power-law spatial distributions for the CR and gas densities, $N_\mathrm{CR}(r) = N_0 r^{-\alpha_1}$ and $n_\mathrm{g}(r) = n_0 r^{-\alpha_2}$,  and noting that for a fixed angular size $\vartheta$, the linear size of the source scales as $R=\vartheta d$, the flux scales as\footnote{For the specific case of $\alpha_1 + \alpha_2 = 2$ (for instance, when both CR and gas densities drop as $1/r$), the flux becomes $F_\gamma \propto \ln(\vartheta d)/d^2$, which would disfavor a distant source.}:

\begin{equation} \label{eq:gamfluxpl}
    F_\gamma \propto N_0 n_0 d^{-(\alpha_1 + \alpha_2) + 1}.
\end{equation}

\noindent Equation \eqref{eq:gamfluxsph} assumes that the volume filled by CRs extends at least to the radius $R$. This imposes a lower limit on the propagation speed --- and correspondingly on the diffusion coefficient --- to ensure that particles could reach the physical distance $R = \vartheta d$ within the operational lifetime of the accelerator.

Crucially, for a fixed accelerator power $L_\mathrm{CR}$ in a similar gas environment, the $\gamma$-ray flux does not follow the standard $1/d^2$ dilution if the cosmic-ray distribution is maintained by continuous injection. From an observational perspective, given that a fixed angular size implies consistent instrumental performance, the association of the Cygnus Bubble with the distant Cygnus X-3 is physically consistent with the available LHAASO data \citep{Cao2025}. Below, we explore this possibility through rigorous calculations that strictly treat particle transport. Specifically, we model the diffusion of particles after their escape from the binary system into the ISM, utilizing a realistic gas distribution surrounding Cygnus X-3 on scales up to \qty{1}{kpc}. This analysis accounts for the non-isotropic nature of the environment, considering propagation both along the Galactic disk and perpendicular to the Galactic plane.

The paper is organized as follows: in Sec. \ref{sec:model}, we describe the analytical and numerical methods used to calculate the cosmic-ray distributions, the resulting $\gamma$-ray spectra, and the surface brightness profiles, based on the relevant assumptions regarding the gas distributions. In Sec. \ref{sec:result}, we present the results for various injection parameters and discuss their physical implications. Finally, we summarize our findings, discuss the results, and provide concluding remarks in Sec. \ref{sec:conclusion}.

\section{The model} \label{sec:model}

After being accelerated, particles (for simplicity, we consider only protons) escape the source and propagate diffusively through the surrounding ISM. Assuming spherical symmetry, with the accelerator located at the center and a homogeneous environment, the transport equation \citep{Aharonian1996,Bosch-Ramon2005} for cosmic rays is

\begin{equation} \label{eq:diffloss}
    \frac{\partial N}{\partial t} - \frac{D}{r^2}\frac{\partial}{\partial r}\left(r^2\frac{\partial N}{\partial r}\right) - \frac{\partial (bN)}{\partial E} = Q,
\end{equation}

\noindent where $N(E, r,t)$ is the differential number density of relativistic particles, $D(E)$ is the spatial diffusion coefficient, $b(E) = -dE/dt$ is the energy-loss rate, and $Q(E,r,t)$ is the source function.

The dominant energy loss process for CR protons is the inelastic collisions with ambient gas. The corresponding cooling time in a pure hydrogen medium is $t_\mathrm{pp} = (\kappa \sigma_\mathrm{inel} c n_\mathrm{H})^{-1} \simeq 5.3\times10^7\,(n_\mathrm{H}/1\,\mathrm{cm}^{-3})^{-1}(\sigma_\mathrm{inel}/40\,\mathrm{mb})^{-1}\,\mathrm{yr}$, where $c$ is the speed of light, $n_\mathrm{H}$ is the number density of ambient hydrogen atoms, $\kappa \simeq 0.5$ is the inelasticity coefficient, and $\sigma_\mathrm{inel}$ the inelastic \textit{pp} collision cross section.

The cross section $\sigma_\mathrm{inel}$ is a weak function of energy and, in the relativistic limit, can be presented, using the parametrization of \citet{Kafexhiu2014}, as $\sigma_\mathrm{inel}(E) = 30.7 - 0.96 \ln x + 0.18(\ln x)^2\,\mathrm{mb}$, with $x = E/T_\mathrm{th}$, where $T_\mathrm{th} \approx \qty{0.2797}{\GeV}$, is the proton’s threshold kinetic energy for pion production.

The spatial diffusion coefficient is parameterized as

\begin{equation} \label{eq:diffcoe}
    D(E) = D_0 \left(\frac{E}{\qty{1}{\PeV}}\right)^\delta,
\end{equation}

\noindent where $D_0$ and $\delta$ depend on the level and spectrum of magnetic turbulence in the ISM \citep{Strong2007}. Since the origin of interstellar turbulence is not fully understood, $D_0$ is treated as a free parameter. We consider both Kolmogorov ($\delta=1/3$) and Iroshnikov-Kraichnan ($\delta=1/2$) turbulence spectra. We do not take into account here that the diffusion coefficient transitions to scale as $E^2$, when the gyro-radii of CRs are larger than the correlation length of interstellar magnetic turbulence \citep{Subedi2017,Giacinti2018,Pezzi2022}.

We assume continuous injection starting at $t=0$, so that

\begin{equation}
    Q(E, r, t) = q(E)\delta^3(\mathbf{r})\theta(t),
\end{equation}

\noindent where $\delta^3(\mathbf{r})$ is the Dirac delta function and $\theta(x)$ is the Heaviside step function. The injection spectrum is taken as

\begin{equation}
    q(E) = Q_0 E^{-s} \exp\left(-\frac{E}{E_0}\right),
\end{equation}

\noindent where $s$ is the spectral index and $E_0$ is the cutoff energy. The total kinetic power injected into cosmic-ray protons is $\dot{W}_\mathrm{CR} = \int_{E_\mathrm{min}}^\infty q(E)EdE$; $E_\mathrm{min}$ is the minimum proton energy. 

We do not explicitly address the underlying acceleration mechanisms of CRs, but note that super-Eddington accreting X-ray binaries may accelerate protons up to several tens of PeV \citep[e.g.,][]{Peretti2025,Wang2025}. Throughout this paper, we fixed the cutoff energy at $E_0 = \qty{10}{\PeV}$.

The solution of Eq. \eqref{eq:diffloss} for the continuous injection is \citep{Syrovatskii1959} 

\begin{equation} \label{eq:sol1}
  N(E,r,t) = \int_0^t \frac{b(E')}{b(E)}q(E')\frac{1}{[4\pi\lambda(E,E')]^{3/2}}\exp\left[-\frac{r^2}{4\lambda(E,E')}\right]dt',
\end{equation}

\noindent where $E'$ is defined through 

\begin{equation}
    t-t'=\int_E^{E'}\frac{dE_1}{b(E_1)},
\end{equation}

\noindent and

\begin{equation}
    \lambda(E,E') = \int_E^{E'}\frac{D(E_1)}{b(E_1)}dE_1.
\end{equation}

\noindent For ultrarelativistic protons, because of the weak energy dependence of  $\sigma_{\mathrm{inel}}$, we have $E'=E\exp[(t-t')/t_\mathrm{pp}]$, thus \citep{Aharonian1996}

\begin{equation}
  \lambda(E,E') = D(E)(t-t') \frac{\exp[\delta(t-t')/t_\mathrm{pp}] - 1}{\delta(t-t')/t_\mathrm{pp}}.
\end{equation}

\noindent In the limit $t\ll t_\mathrm{pp}$, Eq. \eqref{eq:sol1} is reduced to

\begin{equation} \label{eq:sol2}
    N(E,r,t) = \frac{q(E)}{D(E)}\frac{\mathrm{erfc}(r/r_\mathrm{d})}{4\pi r}.
\end{equation}

\noindent Here $\mathrm{erfc}(x)$ is the complementary error function and $r_\mathrm{d} = \sqrt{4D(E)t}$ is the diffusion radius.

\subsection{Applying to Cygnus X-3}

Due to the large distance of Cygnus X-3 and the strong Galactic extinction along its line of sight, a detailed, high-resolution study of the local ISM surrounding the source is currently lacking. To establish a baseline for the $\gamma$-ray production calculations, we first assume a spatially uniform ambient hydrogen density, $n_\mathrm{H}$, and then proceed to a more realistic treatment that accounts for the Galactic disk’s global vertical structure.

Following empirical models of the atomic hydrogen (\textsc{H\,i}) distribution, we adopt a stratified density profile that accounts for the increase of the gas scale height with Galactocentric radius in the outer Galaxy \citep{Kalberla2008,Kalberla2009}. At the Galactocentric radius of Cygnus X-3 ($R \approx \qty{10}{kpc}$), the weakening of the Galactic gravitational potential leads to a vertically extended gas disk with a characteristic scale height of $z_c = \qty{300}{pc}$ \citep{Kalberla2008}.

We model this vertical dependence using a simple exponential profile

\begin{equation} \label{eq:gaszdist}
n_\mathrm{H}(z) = n_{0} \exp \left( -\frac{|z - z_{\mathrm{mid}}|}{z_c} \right)
\end{equation}

\noindent where $n_{0}$ is the mid-plane density and $z_\mathrm{mid}=0$. 

This vertically extended gas distribution provides an extended hadronic target for cosmic rays injected by the Cygnus X-3 jets. In this way, the model bridges the lack of local high-resolution ISM data with the well-established large-scale structure of the Milky Way, enabling a more realistic estimate of the diffuse ultra-high-energy $\gamma$-ray luminosity produced via \textit{pp} interactions in the Cygnus region.

Once the distribution of CRs around the accelerator is obtained, the emissivity per hydrogen atom (in units of \unit{photons.\second^{-1}.\steradian^{-1}.\GeV^{-1}}) of $\gamma$ rays generated from the inelastic \textit{pp} collisions is computed as following \citep{Kafexhiu2014}

\begin{equation}
    q_\gamma(E_\gamma,r,t) = \frac{c}{4\pi}\int_{E_\gamma}^\infty \frac{d\sigma_{\gamma}}{dE_\gamma}(E_\gamma, E) N(E,r,t) dE,
\end{equation}

\noindent where $d\sigma_\gamma/dE_\gamma(E_\gamma, E)$ is the differential production cross section of $\gamma$ rays, which is provided by the package \texttt{aafragpy} \citep{Koldobskiy2021}. Given that the emissivity and gas distribution are known, the $\gamma$-ray intensity is

\begin{equation}
    I_\gamma(E_\gamma, \vartheta, \varphi, t) = \varepsilon_\mathrm{M} \int_{\mathrm{L.O.S.}} n_\mathrm{H}(z) q_\gamma(E_\gamma,r,t)\,dl,
\end{equation}

\noindent where the integration is along the line of sight (L.O.S.), whose direction is defined by the angular separation from the accelerator $\vartheta$ and the azimuthal angle $\varphi$, and $\varepsilon_\mathrm{M}=2.0$ is the nuclear enhancement factor accounting for the contribution to $\gamma$-ray production of nuclei heavier than hydrogen in both CRs and ambient gas \citep{Mori2009,Kachelriess2014}. The nuclear enhancement is calculated assuming that the CR and ISM compositions are the same as the local ones. When the gas distribution is homogeneous, the $\gamma$-ray intensity is independent of $\varphi$. On the other hand, when the gas distribution is nonhomogeneous perpendicular to the Galactic plane, we can define the azimuthally averaged $\gamma$-ray intensity as $(2\pi)^{-1}\int_0^{2\pi} I(E_\gamma, \vartheta, \varphi, t)\,d\varphi$. The $\gamma$-ray flux is

\begin{equation}
  F_\gamma(E_\gamma, t) = \int_0^{\arcsin(R/d)} d\vartheta\,\sin\vartheta \cos\vartheta \int_{0}^{2\pi}d\varphi\, I_\gamma(E_\gamma, \vartheta, \varphi, t),
\end{equation}

\noindent where $d$ is the distance to the accelerator, and $R$ is the dimension of the emission region. We assume that $R=\qty{1}{kpc}$, which corresponds to an angular radius of $\qty{\sim6}{\degree}$ for the emission region, given that the distance to Cygnus X-3 is $d=9.67$ kpc \citep{Reid2023}.

The CRs near their sources can excavates a cavity with a radius of $\sim$10--\qty{50}{pc}, resulting in partial evacuation of gas from the cavity \citep{Schroer2022}. We demonstrate here that this evacuation effect has no significant impact on our estimates of CR acceleration efficiency in Sec. \ref{sec:result}. The diffusion radius of CRs is $r_\mathrm{d}=\sqrt{4Dt_\mathrm{age}}\simeq\qty{600}{pc}$ at \qty{1}{\PeV} with $D=\qty{3e29}{\cm^2\per\s}$ and $t_\mathrm{age}=\qty{100}{kyr}$ (refer to Sec. \ref{sec:result} on the related discussion of model parameters). For the stationary injection, the CR radial distribution is proportional to $1/r$ when $r \ll r_\mathrm{d}$ according to Eq. \eqref{eq:sol2}. Assuming a homogeneous gas distribution, the $\gamma$-ray emissivity is also proportional to $1/r$, then the $\gamma$-ray flux originating from within radius $r$ is $F_\gamma(<r)\propto \int_0^r {r'}^{-1} 4\pi {r'}^2d{r'} =2\pi r^2$, and thus, $F_\gamma(<r_1)/F_\gamma(<r_2) = (r_1/r_2)^2$. Now, taking $r_1=\qty{50}{pc}$ and $r_2=\qty{200}{pc}$, we can see that $F_\gamma(<50\,\mathrm{pc}) / F_\gamma(<200\,\mathrm{pc}) = 1/16$, which means that the $\gamma$-ray flux originating from within \qty{50}{pc} is much smaller than that from within \qty{200}{pc}. Taking the evacuation into account, the contribution to $\gamma$-ray emission from within \qty{50}{pc} is much smaller, and the $\gamma$-ray flux mainly originates from between \qty{50}{pc} and $r_\mathrm{d}$. As a consequence, our estimates on the efficiency are not affected significantly.

\section{results and discussions} \label{sec:result}

Before discussing the implications of our results, we discuss briefly the selection of model parameters. Firstly, we assume that the injection spectral index $s=2.0$ is fixed, which is consistent with the prediction of diffusive shock acceleration theory in strong shocks \citep[e.g., see][]{Malkov2001}. The strong termination shocks driven by fast outflows (winds and jets), which can reach mildly relativistic velocities, in ultra-luminous X-ray sources such as the microquasar Cygnus X-3 \citep{Veledina2024} could accelerate effectively CRs to several tens of PeV \citep[e.g., see][]{Peretti2025,Wang2025}, while other underlying acceleration mechanisms can not be excluded. Secondly, since we do not know exactly how long the injection process sustains, we assume that the elapsed time denoted by $t_\mathrm{age}$ since the injection starts from $t=0$ is a free parameter, and we choose four different values, i.e., $t_\mathrm{age}=$ 100 (solid lines), 200 (dashed lines), 300 (dotdashed lines), 400 (dotted lines) kyr, given that the companion of Cygnus X-3 is a Wolf-Rayet star \citep{Kerkwijk1992,Kerkwijk1996}. Then, we leave only the diffusion coefficient normalization $D_0$, the ambient gas density $n_\mathrm{H}$ ($n_0$) for the homogeneous (nonhomogeneous) gas distribution, and the injection kinetic power $\dot{W}_\mathrm{CR}$ the other three free parameters. The $\gamma$-ray flux and intensity are proportional to the product $n_\mathrm{H}\dot{W}_\mathrm{CR}$, which can be determined by the observed data, once we know $D_0$. As a fiducial value, we assume that $n_\mathrm{H}=\qty{1.0}{cm^{-3}}$ ($n_0=\qty{1.0}{cm^{-3}}$) for the homogeneous (nonhomogeneous) gas distribution. Once choosing $D_0=\qty{3e29}{\cm^2.\s^{-1}}$, we find our model can explain reasonably the observed data for both Iroshnikov-Kraichnan ($\delta=1/2$) and Kolmogorov ($\delta=1/3$) turbulence, by tuning $\dot{W}_\mathrm{CR}$.

The radial profile of $\gamma$-ray intensity can give crucial information on the spatial distribution of CRs and their injection history. To obtain this radial flux distribution, we extracted the photon distribution within the \qty{6}{\degree}-radius region from Figure 1 of \citet{Cao2024}, which contained a total of 66 photon-like events with energies exceeding 400 TeV, with an estimated CR background of 9.5. Using the energy spectrum from Figure 3 of \citet{Cao2024}, we calculated the integrated energy above 400 TeV within the same \qty{6}{\degree}-radius region. After subtracting the CR background, we converted the photon distribution into a radial flux distribution for energies above 400 TeV.

Meanwhile, we estimated the diffuse $\gamma$-ray emission utilizing the proton and helium spectra measured by DAMPE \citep{dampe_proton, dampe_helium} and LHAASO \citep{LHAASO_proton, LHAASO_helium}. For the gas distribution, we employed the HI4PI survey data \citep{HI4PI_2016A&A...594A.116H} and the CfA $^{12}$CO data \citep{CFA_12CO_1987ApJ...322..706D}. The column density of neutral hydrogen was calculated by integrating over the entire velocity range using \citep{wilson2013}:
\begin{equation}
    N_{\rm HI} = 1.8 \times 10^{18} \int T_{\rm mb, HI} \, dV
\end{equation}
The molecular hydrogen column density was derived by:
\begin{equation}
    N_{\rm H_2} = X \, W_{^{12}\rm CO} = X \int T_{\rm mb, ^{12}\rm CO} \, dV
\end{equation}
Here, we adopted a mean CO-to-H$_2$ conversion factor of $X = 2.0 \times 10^{20}~\rm cm^{-2}~K^{-1}~km~s$\citep{The_CO_H2_Conversion_Factor_2013ARA&A..51..207B}. The total hydrogen column density is $N_{\rm H} = N_{\rm HI} + 2N_{\rm H_2}$, and we assumed a helium abundance of $N_{\rm He} = 0.1 N_{\rm H}$ in the ISM. We then used \texttt{aafragpy} \citep{Koldobskiy2021} to calculate the diffuse $\gamma$-ray emission expected from the cosmic ray proton and helium spectra.

\begin{figure}
    \centering
    \includegraphics[width=0.49\textwidth]{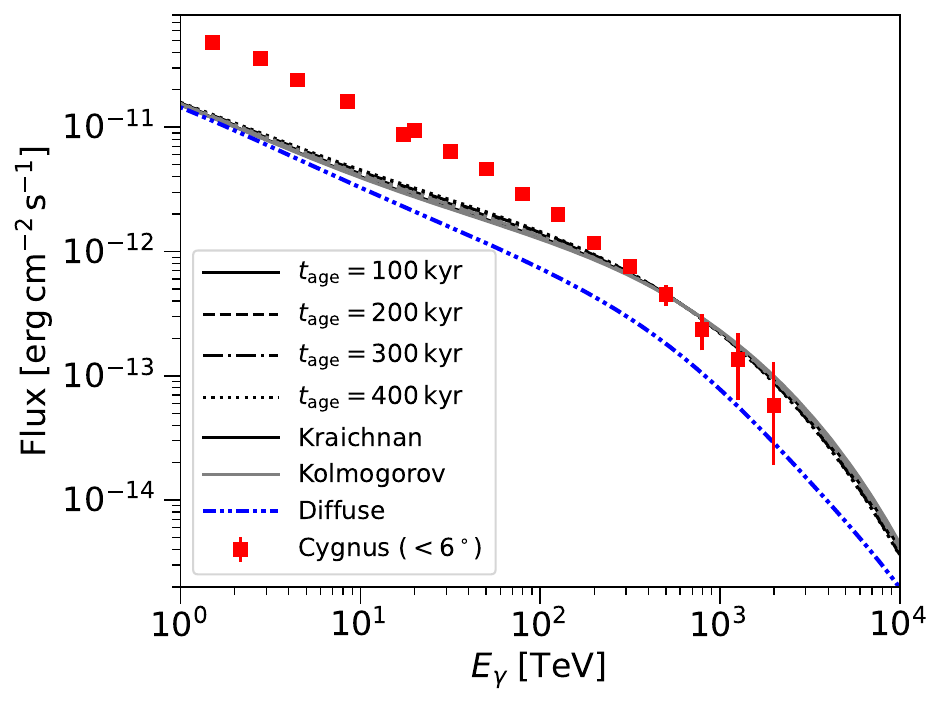}
    \includegraphics[width=0.49\textwidth]{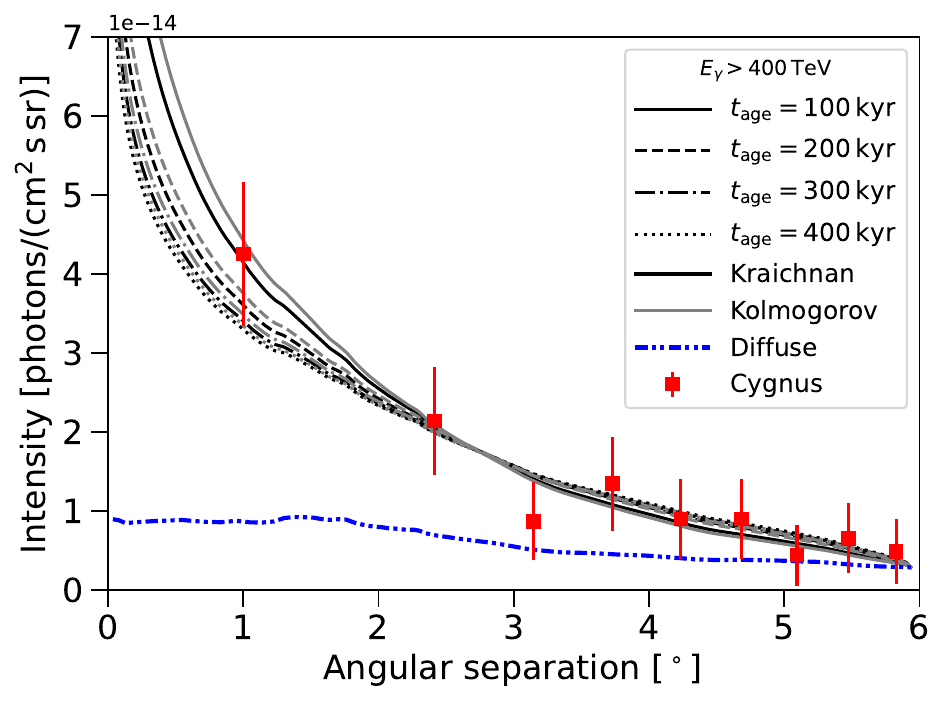}
    \caption{\textbf{Left panel}: the $\gamma$-ray flux from within a \qty{6}{\degree}-radius emission region for Iroshnikov-Kraichnan ($\delta=1/2$; black lines) and Kolmogorov ($\delta=1/3$; grey lines) turbulence phenomenology, when $t_\mathrm{age}=100$ (solid lines), 200 (dashed lines), 300 (dotdashed lines), and 400 (dotted lines) kyr, which is the elapsed time since the injection starts. The blue dashdotdotted line shows the diffuse Galactic $\gamma$-ray flux, which is produced by the cosmic ray ``sea" as measured locally on the Earth, while the black and grey lines show the summation of our model and diffuse $\gamma$-ray fluxes. The red flux points for a \qty{6}{\degree}-radius bubble is taken from \citet{Cao2024}. The diffusion coefficient normalization $D_0=\qty{3e29}{\cm^2.\s^{-1}}$ for both Iroshnikov-Kraichnan and Kolmogorov turbulence, the CR injection spectral index $s=2.0$ and its cutoff energy $E_0=\qty{10}{\PeV}$, the gas distribution is homogeneous and its density $n_\mathrm{H}=\qty{1}{\cm^{-3}}$, while the CR injection kinetic powers are given in Table \ref{tab:crpower}. \textbf{Right panel}: the integrated $\gamma$-ray intensity above 400 TeV. The line styles and corresponding parameters are the same as the left panel.}
    \label{fig:gamflux}
\end{figure}

Fig. \ref{fig:gamflux} shows the fits of our model to the flux of $\gamma$ rays with energies larger than 400 TeV from the Cygnus bubble within a radius of \qty{6}{\degree} as observed by LHAASO \citep{Cao2024} and the radial profile of integrated $\gamma$-ray intensity above 400 TeV (i.e., $\int_{\qty{400}{\TeV}}^\infty I_\gamma(\vartheta, E_\gamma, t)dE_\gamma$) in left and right panel, respectively, for a homogeneous gas distribution. In the left panel, the blue dashdotdotted line shows the diffuse Galactic $\gamma$-ray flux within a \qty{6}{\degree}-radius region around Cygnus X-3, while the black and gray lines show the summation of our model and diffuse $\gamma$-ray fluxes for Iroshnikov-Kraichnan and Kolmogorov turbulence, respectively, when $t_\mathrm{age}=100$ (solid lines), 200 (dashed lines), 300 (dotdashed lines), and 400 (dotted lines) kyr. Moreover, in the right panel, the corresponding lines show the corresponding radial profiles of integrated $\gamma$-ray intensity above \qty{400}{\TeV}. As shown in Fig. \ref{fig:gamflux}, our model can explain reasonably the observed $\gamma$-ray flux and intensity radial profile for energies above 400 TeV. In fact, given the parameters we chose as discussed above, the radial profile of integrated $\gamma$-ray intensity can be fitted reasonably without fine-tuning the parameters, once we fit our model to the observed $\gamma$-ray flux by tuning $\dot{W}_\mathrm{CR}$, whose best-fit values are listed in Table \ref{tab:crpower}. Moreover, we find that the required injection kinetic power $\dot{W}_\mathrm{CR}$ decreases when $t_\mathrm{age}$ increases, as expected. However, when $t_\mathrm{age}>\qty{400}{kyr}$, we find $\dot{W}_\mathrm{CR}$ no longer decreases apparently, when increasing $t_\mathrm{age}$ further. Due to the interplay of continuous injection into and escaping of CRs from the emission region, their distribution tends to $N(E,r,t)=q(E)/[4\pi rD(E)]$ when $r\ll r_\mathrm{d}$, which is satisfied as $t_\mathrm{age}$ is large enough. Given that the kinetic luminosity of Cygnus X-3 is estimated to be $L_\mathrm{kin}=\qty{5e39}{\ergs}$ \citep{Veledina2024,Wang2025}, the required acceleration efficiency ($=\dot{W}_\mathrm{CR}/L_\mathrm{kin}$) is 0.7--\qty{1.6}{\percent} according to Table \ref{tab:crpower}. Even though we adopt a lower gas density, for instance $n_\mathrm{H}=\qty{0.1}{\cm^{-3}}$, the required acceleration efficiency is 7--\qty{16}{\percent}, thus Cygnus X-3 has enough energy budget for accelerating CRs to about \qty{10}{\PeV}.

\begin{figure}
    \centering
    \includegraphics[width=0.49\textwidth]{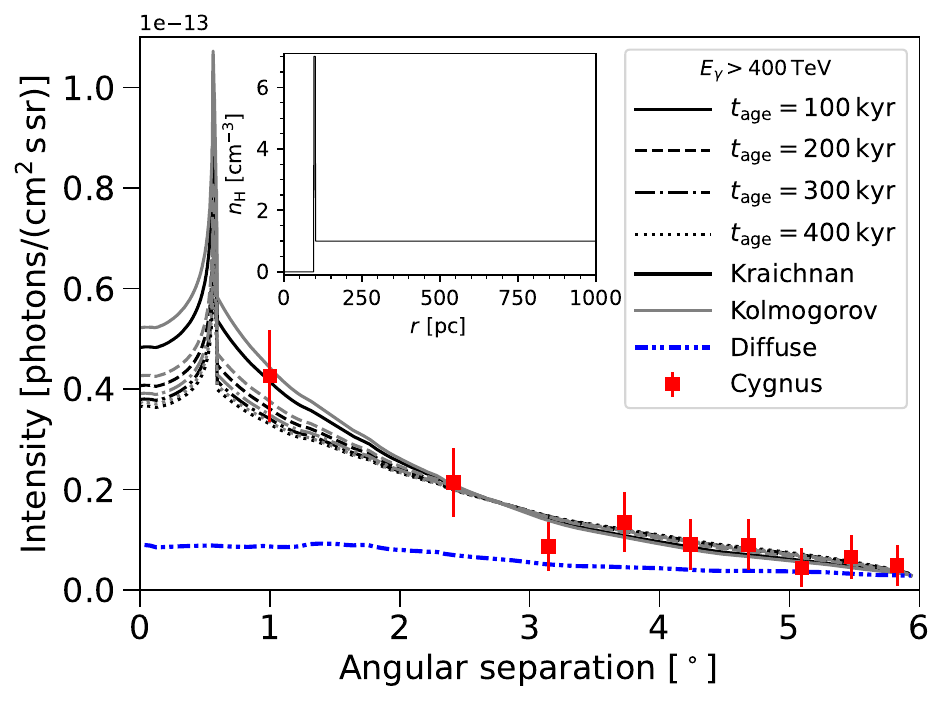}
    \caption{Same as the right panel of Fig. \ref{fig:gamflux}, but the gas density is given as shown in the inset plot.}
    \label{fig:gamprofcav}
\end{figure}

As we have discussed at the end of Sec. \ref{sec:model}, the cavity excavated by the cosmic ray pressure \citep{Schroer2022} has no significant impact on the $\gamma$-ray flux for an emission region with a size of \qty{1000}{pc}. However, such an low-density cavity embedded in the emission region can leave an imprint on the $\gamma$-ray intensity profile. Here, we briefly discuss the aftermath due to the evacuation effect. We assume the cavity has a radius $R_\mathrm{cavity}=\qty{100}{pc}$, and the evacuated gas accumulates in a thin shell with a thickness $\Delta R=\qty{5}{pc}$ located at the outer surface of the cavity. Therefore, the gas density in the shell is $n_\mathrm{shell}=n_\mathrm{H}/[1 - (1 - \Delta R/R_\mathrm{cavity})^3]=7.0n_\mathrm{H}$, where $n_\mathrm{H}=\qty{1}{\cm^{-3}}$ is the gas density outside of the cavity, which we assume is not affected by the evacuation effect. Fig. \ref{fig:gamprofcav} shows the integrated $\gamma$-ray intensity radial profile above \qty{400}{TeV} taking the evacuation effect into account. All parameters are the same as those discussed in the previous paragraph, except the gas distribution. As shown in Fig. \ref{fig:gamprofcav}, there is a characteristic peak around $\vartheta=\qty{0.6}{\degree}$ ($=R_\mathrm{cavity}/d$) in the intensity profile, while we verify that the $\gamma$-ray flux is almost not affected at least for energies above \qty{1}{TeV} for each set of parameters (not shown). However, such a feature can not be resolved by the present LHAASO data, and the future high angular resolution observations may reveal if such a feature exists. Hereafter, we will not consider the evacuation effect.

\begin{figure}
    \centering
    \includegraphics[width=0.49\textwidth]{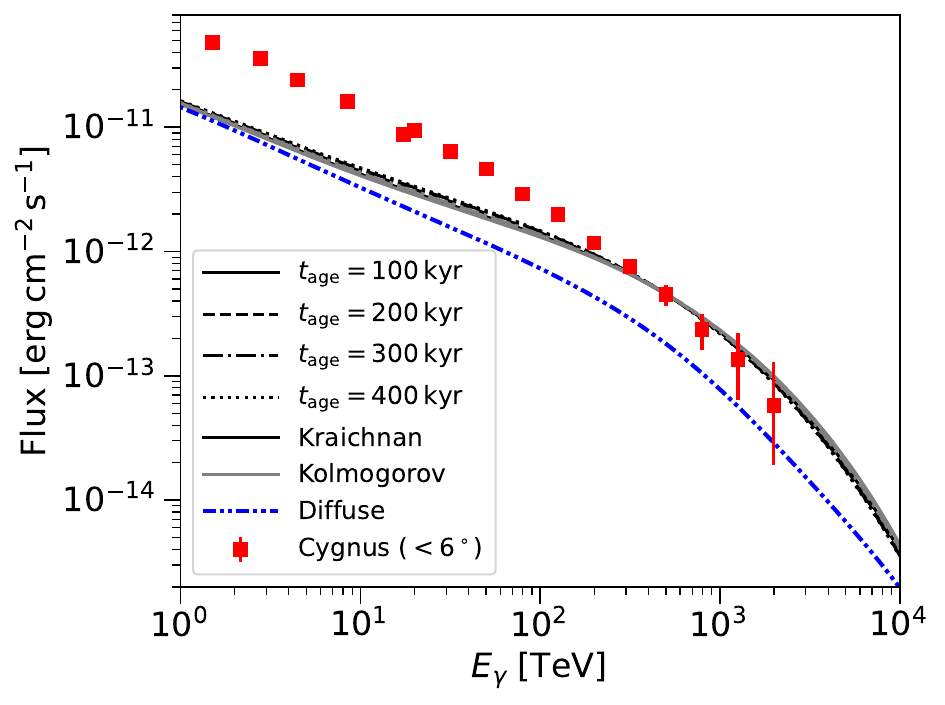}
    \includegraphics[width=0.49\textwidth]{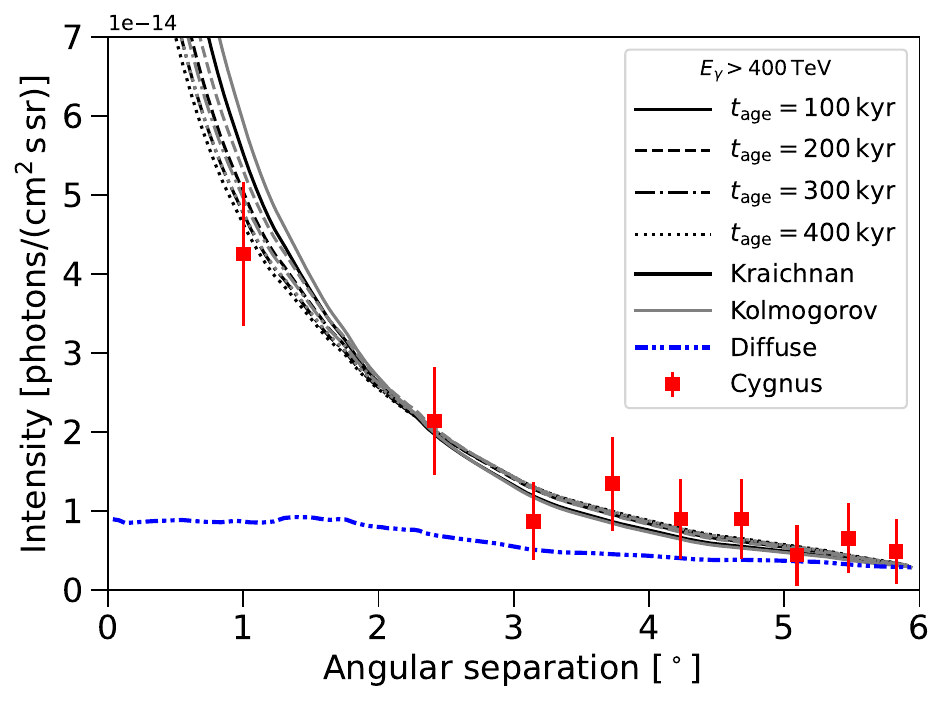}
    \caption{Same as Fig. \ref{fig:gamflux} but the right panel displays the azimuthally averaged integrated $\gamma$-ray intensity above \qty{400}{\TeV}, and the gas distribution is nonhomogeneous and is given by Eq. \eqref{eq:gaszdist} with $n_0=\qty{1}{\cm^{-3}}$ and $z_c=\qty{300}{pc}$, while the CR injection kinetic powers are given in Table \ref{tab:crpower2}.}
    \label{fig:gamfluxhg300}
\end{figure}

While a homogeneous gas distribution is not realistic and should only be regarded as an average over the emission region, we also consider a more realistic nonhomogeneous gas distribution, which has a finite scale height vertical to the Galactic plane as prescribed by Eq. \eqref{eq:gaszdist}. We assume that the mid-plane gas density $n_0=\qty{1.0}{\cm^{-3}}$ and the scale height $z_c=\qty{300}{pc}$. For such a gas distribution, the average gas density within the emission region is about \qty{0.38}{\cm^{-3}}. In Fig. \ref{fig:gamfluxhg300}, the left and right panel show the fits of our model to observed $\gamma$-ray flux and intensity radial profile, respectively, while the best-fit values of $\dot{W}_\mathrm{CR}$ are listed in Table \ref{tab:crpower2}, assuming $D_0=\qty{3e29}{\cm^2.\s^{-1}}$ for both Iroshnikov-Kraichnan and Kolmogorov turbulence. Similar to the case for a homogeneous gas distribution, our model for the nonhomogeneous gas distribution can also explain reasonably the observed $\gamma$-ray flux and intensity radial profile for energies above \qty{400}{\TeV}. According to Table \ref{tab:crpower2}, the required acceleration efficiency is 1.6--\qty{3.2}{\percent}, for the nominal kinetic luminosity $L_\mathrm{kin}=\qty{5e39}{\ergs}$ of Cygnus X-3. Therefore, our results suggest that the $\gamma$ rays with energies above 400 TeV from Cygnus bubble observed by LHAASO \citep{Cao2024} may originate from the $\mu$QSH forming around the super-PeVatron microquasar Cygnus X-3, similar to the other five Galactic microquasars reported recently by LHAASO \citep{Liu2025}.

The spatial diffusion coefficient plays a vital role in determining the spatial distribution of CRs, and hence the resulting $\gamma$-ray morphology. Therefore, we discuss it further. The spatial diffusion coefficient in the ISM is $D(E=\qty{1}{GeV}) \sim \qty{3e28}{\cm^2.\s^{-1}}$, based on the investigations on the
propagation of CRs in the Galaxy and on the diffuse Galactic $\gamma$-ray emission \citep{Strong2007}. When extrapolating the empirical diffusion coefficient to $E=\qty{1}{\PeV}$, $D(E=\qty{1}{\PeV}) \sim \qty{3e31}{\cm^2.\s^{-1}}$ (\qty{3e30}{\cm^2.\s^{-1}}) for Iroshnikov-Kraichnan (Kolmogorov) turbulence, which is two (one) orders of magnitude larger than $D_0=\qty{3e29}{\cm^2.\s^{-1}}$ we obtained. Although self-generated turbulence from CR streaming instability can suppress diffusivity around their sources, this suppression is confined to a region of a few tens of parsecs \citep{Schroer2022}. On the other hand, strong extrinsic turbulence could be injected by the relativistic jets of Cygnus X-3, producing an extended region of suppressed diffusivity. The spatial diffusion coefficient of CRs according to quasi-linear theory \citep{Schlickeiser1989} for $r_\mathrm{g} < l_\mathrm{c}$ is 

\begin{equation} \label{eq:qltdiff} 
  D(p) \simeq \frac{\beta cr_\mathrm{g}}{3} \left(\frac{l_\mathrm{c}}{r_\mathrm{g}}\right)^{1 - \delta} \frac{1}{\eta} = \begin{cases} \qty{1.85e28}{\cm^2.\s^{-1}} \beta \left(\frac{pc}{\unit{\PeV}}\right)^{1/2} \left(\frac{B}{\qty{3}{\micro G}}\right)^{-1/2} \left(\frac{l_\mathrm{c}}{\unit{pc}}\right)^{1/2}\eta^{-1}, & \delta = 1/2, \\ \qty{2.19e28}{\cm^2.\s^{-1}} \beta \left(\frac{pc}{\unit{\PeV}}\right)^{1/3} \left(\frac{B}{\qty{3}{\micro G}}\right)^{-1/3} \left(\frac{l_\mathrm{c}}{\unit{pc}}\right)^{2/3}\eta^{-1}, & \delta = 1/3, \end{cases}
\end{equation}

\noindent where $c$ is the speed of light, $\beta$ the ratio of CR velocity to $c$, $r_\mathrm{g}=pc/eB = \qty{0.36}{pc}\,(pc/\unit{\PeV}) (B/\qty{3}{\micro G})^{-1}$ the gyro-radius of CRs, $l_\mathrm{c}$ the correlation length of magnetic turbulence, and $\eta=(\delta B/B)^2$ the magnetic turbulence level. For $r_\mathrm{g} > l_\mathrm{c}$, $D(p)\propto \beta p^2$ \citep{Subedi2017,Giacinti2018,Pezzi2022}. The typical correlation length of magnetic turbulence in the interarm regions is $\qty{10}{pc} \lesssim l_\mathrm{c} \lesssim \qty{100}{pc}$ \citep{Haverkorn2008,Hollins2017}. In the spiral arms, stellar sources dominate the energy injection for the turbulence cascade, and the typical correlation length is $l_\mathrm{c}\sim \qty{1}{pc}$ \citep{Haverkorn2008}. According to \citet{Reid2023}, Cygnus X-3 is located in the Outer spiral arm. If Cygnus X-3 dominates the turbulence energy injection for the emission region with size $R=\qty{1000}{pc}$, and if we assume $l_\mathrm{c}\simeq \qty{1}{pc}$, then according to Eq. \eqref{eq:qltdiff}, the turbulence level should be $\eta \sim 1/10$ for $D_0 = \qty{3e29}{\cm^2.\s^{-1}}$, assuming the magnetic field strength $B=\qty{3}{\micro G}$. In such a situation, however, Eq. \eqref{eq:qltdiff} is not applicable for $p>\qty{1}{\PeV/\clight}$. If we assume a larger correlation length $l_\mathrm{c}\simeq \qty{10}{pc}$, Eq. \eqref{eq:qltdiff} is applicable up to $p=\qty{10}{\PeV/\clight}$, and the required turbulence level is $\eta \sim 1/3$. Though there are still many uncertainties in our understanding of the interaction between PeV CRs and magnetic turbulence \citep{Hu2026}, the transition of the scattering regime should be incorporated into the modeling of PeV CR propagation, which is not taken into account in the present work.

\begin{table}[ht]
    \centering
    \begin{threeparttable}
        \caption{The CR injection kinetic power for a homogeneous gas distribution.}
        \label{tab:crpower}
        \vspace{2mm}
        \begin{tabular}{ccc}
            \toprule
            $t_\mathrm{age}$ & \multicolumn{2}{c}{$\dot{W}_\mathrm{CR}$ [\qty{e37}{\ergs}]} \\
            \cmidrule(lr){2-3}
            [kyr] & Kraichnan\tnote{a} & Kolmogorov\tnote{a} \\ 
            \midrule
            100 & 8.0 & 7.2 \\
            200 & 5.6 & 4.8 \\
            300 & 4.8 & 4.0 \\
            400 & 4.4 & 3.6 \\
            \bottomrule
        \end{tabular}
        \begin{tablenotes}
            \small
            \item[a] The CR injection spectrum is an exponentially cutoff power-law function with the spectral index $s=2.0$ and the cutoff energy $E_0=\qty{10}{\PeV}$, while the minimum proton kinetic energy is $E_\mathrm{min}=\qty{1}{\GeV}$. We have assumed that the gas distribution is homogeneous and its density is $n_\mathrm{H}=\qty{1}{\cm^{-3}}$. The diffusion coefficient normalization $D_0=\qty{3e29}{\cm^2.\s^{-1}}$ for both Iroshnikov-Kraichnan and Kolmogorov turbulence.
        \end{tablenotes}
    \end{threeparttable}
\end{table}

\begin{table}[ht]
    \centering
    \begin{threeparttable}
        \caption{The CR injection kinetic power for a nonhomogeneous gas distribution.}
        \label{tab:crpower2}
        \vspace{2mm}
        \begin{tabular}{ccc}
            \toprule
            $t_\mathrm{age}$ & \multicolumn{2}{c}{$\dot{W}_\mathrm{CR}$ [\qty{e37}{\ergs}]} \\
            \cmidrule(lr){2-3}
            [kyr] & Kraichnan\tnote{a} & Kolmogorov\tnote{a} \\ 
            \midrule
            100 & 16.0 & 14.4 \\
            200 & 12.0 & 10.4 \\
            300 & 10.4 & 8.8 \\
            400 & 9.6 & 8.0 \\
            \bottomrule
        \end{tabular}
        \begin{tablenotes}
            \small
            \item[a] The CR injection spectrum is an exponentially cutoff power-law function with the spectral index $s=2.0$ and the cutoff energy $E_0=\qty{10}{\PeV}$, while the minimum proton kinetic energy is $E_\mathrm{min}=\qty{1}{\GeV}$. We have assumed that the gas distribution is nonhomogeneous and is given by Eq. \eqref{eq:gaszdist} with $n_0=\qty{1}{\cm^{-3}}$ and $z_c=\qty{300}{pc}$. The diffusion coefficient normalization $D_0=\qty{3e29}{\cm^2.\s^{-1}}$ for both Iroshnikov-Kraichnan and Kolmogorov turbulence.
        \end{tablenotes}
    \end{threeparttable}
\end{table}


\begin{figure}
    \centering
    \includegraphics[width=0.49\textwidth]{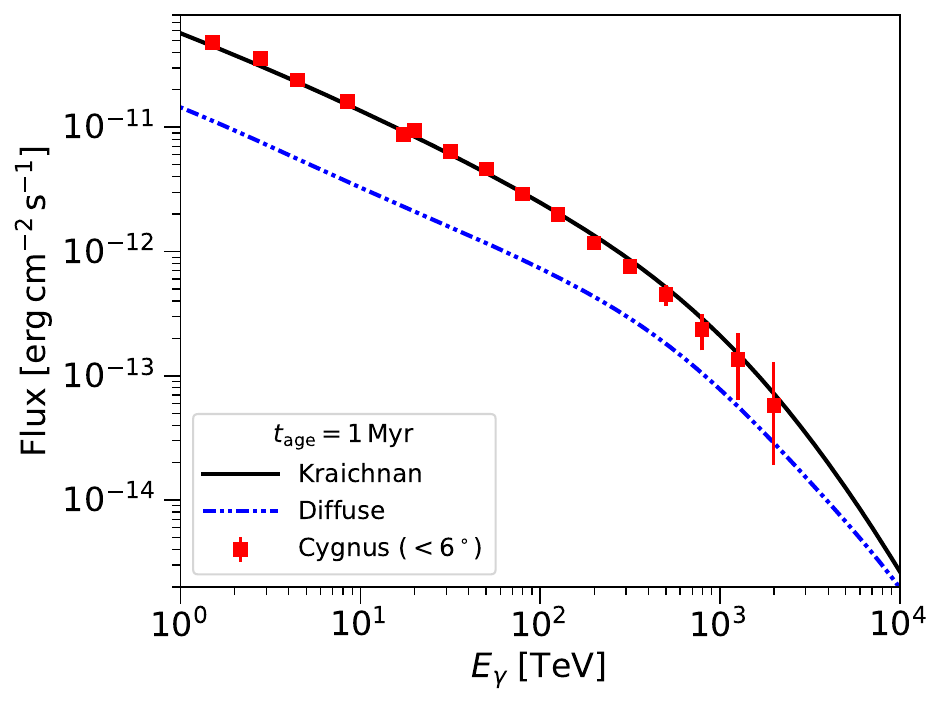}
    \caption{The $\gamma$-ray flux from within a \qty{6}{\degree}-radius emission region for Iroshnikov-Kraichnan turbulence ($\delta=1/2$), assuming that $t_\mathrm{age}=\qty{1}{Myr}$. The black line shows the summation of our model and diffuse Galactic $\gamma$-ray fluxes, while the latter is shown by the blue dashdotdotted line. The red flux points for a \qty{6}{\degree}-radius bubble is taken from \citet{Cao2024}. The diffusion coefficient normalization $D_0=\qty{3e29}{\cm^2.\s^{-1}}$, the CR injection spectral index $s=2.45$ and its cutoff energy $E_0=\qty{10}{\PeV}$, the gas distribution is nonhomogeneous and is given by Eq. \eqref{eq:gaszdist} with $n_0=\qty{1}{\cm^{-3}}$ and $z_c=\qty{300}{pc}$, and the CR injection kinetic power above \qty{1}{\TeV} is $\dot{W}_\mathrm{CR}(>\qty{1}{\TeV})=\qty{5.6e38}{erg/s}$.}
    \label{fig:fluxsoftcr}
\end{figure}

Finally, we briefly discuss the possibility if Cygnus X-3 can account for the $\gamma$-ray flux of the \qty{6}{\degree}-radius Cygnus bubble as observed by LHAASO, not limiting to energies above \qty{400}{\TeV}. In order to account for the observed wideband $\gamma$-ray flux from \qty{1}{\TeV} to \qty{2}{\PeV}, a soft CR injection spectrum above \qty{1}{\TeV} is needed. Here, we assume an Iroshnikov-Kraichnan turbulence spectrum, i.e., $\delta=1/2$, and the diffusion coefficient normalization $D_0=\qty{3e29}{\cm^2.s^{-1}}$ as obtained previously. Furthermore, we assume the CR injection duration $t_\mathrm{age}=\qty{1}{Myr}$. In order to fit to the observed flux, as shown in Fig. \ref{fig:fluxsoftcr}, the CR injection spectral index we obtained is $s=2.45$, assuming an exponential cutoff energy $E_0=\qty{10}{\PeV}$. The required CR injection kinetic power above \qty{1}{TeV} is $\dot{W}_\mathrm{CR}(>\qty{1}{\TeV})=\qty{5.6e38}{erg/s}$, which is about \qty{11}{\percent} of the kinetic luminosity of Cygnus X-3. If we extrapolate the CR injection spectrum to \qty{1}{\GeV}, then the required CR injection kinetic power above $\qty{1}{\GeV}$ is higher than the kinetic luminosity of Cygnus X-3. However, a harder CR injection spectrum below \qty{1}{TeV} can not be excluded, thus the LHAASO observation can not constrain the low-energy spectrum.

\section{Conclusion} \label{sec:conclusion}
In this work, we proposed a simple propagation model, which is based on the diffusion equation \eqref{eq:diffloss}, to explain the origin of $\gamma$ rays with energies above \qty{400}{\TeV} coming from the direction of Cygnus bubble reported recently by LHAASO \citep{Cao2024} within a region with a radius of \qty{6}{\degree}. With only a few free parameters, our model can explain reasonably the observed $\gamma$-ray flux and integrated intensity radial profile above \qty{400}{\TeV} within the \qty{6}{\degree}-radius region, assuming that CRs are injected continuously by the microquasar Cygnus X-3 for a duration of {$\sim$}\qty{100}{kyr} with an energy spectral index $s=2.0$ and an exponential cutoff energy $E_0=\qty{10}{\PeV}$. The CR spatial diffusion coefficient at $E=\qty{1}{\PeV}$ we obtained is $D_0=\qty{3e29}{\cm^2.\s^{-1}}$ for both Iroshnikov-Kraichnan and Kolmogorov turbulence, which is a plausible value in accordance with the quasilinear theory, though the transition of the scattering regime in the magnetic turbulence is not taken into account. Given that the kinetic luminosity of Cygnus X-3 is about $L_\mathrm{kin}=\qty{5e39}{\ergs}$ \citep{Veledina2024,Wang2025}, our results imply that the CR acceleration efficiency is 0.7--\qty{1.6}{\percent} (1.6--\qty{3.2}{\percent}) for the homogeneous (nonhomogeneous) gas distribution assuming a density $n_\mathrm{H}=\qty{1.0}{\cm^{-3}}$ ($n_0=\qty{1.0}{\cm^{-3}}$ with a scale height of \qty{300}{pc}). Thus, our results suggest that Cygnus X-3, which is capable of accelerating CRs to {$\sim$}\qty{10}{\PeV}, can explain UHE photons above \qty{400}{\PeV} coming from the direction of Cygnus bubble within a \qty{6}{\degree}-radius region, despite its large distance ($d=\qty{9.67}{kpc}$) from the Earth. This scenario presents a unique case where we simultaneously detect both the primary accelerator, Cygnus X-3, and the surrounding ``cosmic-ray halo" formed by the historical accumulation of particles injected by this source into the ISM.

While the UHE emission above $400~\rm TeV$ is the main focus of this work, the lower-energy (GeV–TeV) emission of Cygnus bubble can be treated as foreground radiation from an extended gamma-ray source linked to the Cygnus OB2 association. In other microquasars, such as SS 433, V4641 Sgr, and GRS 1915+105, the $\gamma$-ray observations have shown that acceleration happens further out in the jet termination shocks \citep{Liu2025,HAWC2018,HESS2024,HAWC2024,HESS2026}. Our results demonstrate that Cygnus X-3 is among these objects. In this case, Cygnus X-3 would be a ``dual" source: the orbitally modulated PeV photons come from the very compact inner region, while the more extended jet termination regions can be responsible for multi-TeV CRs. Instead of just a stellar-wind cavity, the Cygnus bubble at energies above \qty{400}{\TeV} can be seen as a massive microquasar nebula.

The future deployment of next-generation Imaging Atmospheric Cherenkov Telescopes (IACTs), such as the Cherenkov Telescope Array (CTA) \citep{CTA2019}, the Astrofisica con Specchi a Tecnologia Replicante Italiana (ASTRI) \citep{ASTRI2022}, and the proposed Large Array of Cherenkov Telescopes (LACT) \citep{Zhang2026}, will be instrumental in validating the Cygnus X-3 injector hypothesis by providing superior angular resolution ($<0.05^\circ$) in the UHE band. While LHAASO has effectively identified the Cygnus region as a Super PeVatron, its current resolution remains insufficient to fully disentangle the complex line-of-sight superposition between the foreground Cygnus Cocoon and the background Cygnus X-3 environment. IACTs will allow for the spatial resolution of a compact, point-like core at the microquasar’s coordinates and the mapping of energy-dependent morphology, where a shrinking emission size at higher energies would serve as a classic signature of a discrete injector.


\begin{acknowledgments}

Rui-zhi Yang is supported by the National Natural Science Foundation
of China under grants 12393854, 12588101, and by the natural science funding of Sichuan Province under grant 2025ZNSFSC0065. Rui-zhi Yang gratefully acknowledge the support of Cyrus Chun Ying Tang Foundations and of the studio of Academician Zhao Zhengguo, Deep Space Exploration Laboratory. F. Aharonian acknowledges support from Sichuan Province Science and Technology Department.

\end{acknowledgments}

\begin{contribution}

All authors contributed equally.


\end{contribution}

\bibliography{main}{}
\bibliographystyle{aasjournalv7}



\end{document}